\documentclass[aps,pre,reprint]{revtex4-1}
\usepackage{times}		% this uses fonts which will look nice in PDF format
\usepackage{booktabs}

\usepackage{amssymb,amstext,amsmath, epsf}
\usepackage{graphicx,graphics,epsfig,wrapfig}
\usepackage{listings}
\usepackage{float}
\usepackage{esint}

\usepackage{color}
\usepackage[english]{babel}
\usepackage[T1]{fontenc}
\usepackage[utf8]{inputenc}

\def\apj{{ApJ}}
\def\apjl{{ApJL}}

\def\mnras{{MNRAS}}

\def\prl{{PRL}}

\def\pre{{PRE}}

\def\jgr{{Journal of Geophysical Research}}
\def\apss{{Astrophysics and Space Science}}

\begin{document}

\title{On Particle Transport and Radiation Production in Sub-Larmor-Scale Electromagnetic Turbulence}
\author{Brett D. Keenan}
\email{bdkeenan@ku.edu}
\affiliation{Department of Physics and Astronomy, University of Kansas, Lawrence, KS 66045}
\author{Mikhail V. Medvedev}
\altaffiliation{Also at the Department of Physics and Astronomy, University of Kansas, Lawrence, KS 66045}
\altaffiliation{Also at the ITP, NRC ``Kurchatov Institute", Moscow 123182, Russia}
\affiliation{Institute for Theory and Computation, Harvard University, 60 Garden St., Cambridge, MA 02138}

%\date{\today}

\begin{abstract}
The relation of particle transport of relativistic particles in plasmas with high-amplitude isotropic sub-Larmor-scale magnetic turbulence to the spectra of radiation simultaneously produced by these particles is investigated both analytically and numerically. We have found that in the asymptotic regime of very small particle deflections the pitch angle diffusion coefficient is directly related to the spectrum of the emitted radiation. Moreover, this spectrum provides much information about the statistical properties of the underlying magnetic turbulence. The transition from small- to large-scale jitter to synchrotron radiation regimes as a function of turbulence properties has also been explored. These results can readily be used to diagnose laboratory and astrophysical plasmas.
\end{abstract}

\maketitle

\section{Introduction}
High-amplitude sub-Larmor-scale electromagnetic turbulence is ubiquitous in high-energy density environments, such as laboratory plasmas produced by high-intensity lasers, e.g., National Ignition Facility, Omega, Hercules, Trident, and others \citep{ren04, huntington12, mondal12, tatarakis03}, and in astrophysical and space plasmas, e.g., at high-Mach-number collisionless shocks in weakly magnetized plasmas \citep{medvedev09, frederiksen04, nishikawa03}, upstream regions of quasi-parallel shocks \citep{sironi06, plotnikov12}, sites of magnetic reconnection \citep{swisdak08, liu09} and others. Studies of plasmas and turbulence in these environments are important for fusion energy sciences and the inertial confinement concept \citep{ren04, tatarakis03}, in particular, as well as to numerous astrophysical systems such as gamma-ray bursts \citep{medvedev09b, medvedev06, reynolds12}, supernovae blast waves \citep{fan11}, jets of quasars and active galactic nuclei \citep{mao07}, shocks in the interplanetary medium \citep{gurnett79}, solar flares \citep{mcateer09} and many more. Such turbulence can be of various origin and thus have rather different properties, from being purely magnetic (Weibel) turbulence \citep{weibel59, medvedev09c}, to various types of electromagnetic turbulence (for example, whistler wave turbulence or turbulence produced by filamentation/mixed mode instability), to purely electrostatic Langmuir turbulence \citep{lemoine09, treumann97, bret05}.

Despite substantial differences, these turbulences share one thing in common: the electromagnetic fields vary on scales much smaller than the characteristic curvature radius of the particles' paths (i.e., the Larmor radius in most cases), which is typically of the order of the plasma inertial length (skin depth), so that the particle orbits are never a well-defined Larmor circle. If the electromagnetic fields are random, which is usually the case of turbulence because of the random phases of fluctuations, their paths diffusively diverge due to pitch-angle diffusion. Radiation simultaneously produced by these particles is neither cyclotron nor synchrotron (for non-relativistic or relativistic particles, respectively) but, instead, carries information about the spectrum of turbulent fluctuations. In this paper we explore the relation of the transport of relativistic particles in isotropic three-dimensional magnetic turbulence and the radiation spectra simultaneously produced by these particles. 

We have found that in the small-angle-deflection regime, the radiation spectrum agrees with the jitter radiation prediction \citep{medvedev00,medvedev06,medvedev11,RK10,TT11}. We also demonstrate that the pitch-angle diffusion coefficient is directly related to and can readily be deduced from spectra of the emitted radiation. This provides a unique way to remotely diagnose high-energy-density plasmas, both in laboratory experiments and in astrophysical systems. This can be particular interest, for example, for the physics of collisionless shocks and Fermi particle acceleration in them, because the latter is intimately related to the diffusion coefficient in the upstream and downstream regions of such shocks.

The rest of the paper is organized as follows. Section \ref{s:analytic} presents the analytic theory. Sections \ref{s:model} and \ref{s:results} describe the numerical techniques employed and the obtained simulation results. Section \ref{s:concl} is the conclusions.

%\section{Pitch-Angle Diffusion and Radiation Production in Small-Scale Magnetic Turbulence}
\section{Analytic theory}
\label{s:analytic}

Let us consider a relativistic electron moving through a non-uniform, inhomogeneous, random, small-scale magnetic field (and we assume that this magnetic micro-turbulence  is statistically homogeneous and isotropic). Because of the random Lorentz force on the electron, it's acceleration and velocity vectors vary stochastically, leading to a random (diffusive) trajectory. We define the field turbulence to be ``small-scale'' when the electron's Larmor radius, $\rho_e = \gamma\beta m_e c^2/e B_\perp$ (where $\gamma$ is the electron's Lorentz factor, $\beta=v/c$ is the dimensionless particle velocity, $m_e$ is the electron mass, $c$ is the speed of light, $e$ is the electric charge, and $B_\perp$ is the component of the magnetic field perpendicular to the electron's velocity vector)  is greater than, or comparable to, the characteristic correlation scale of the magnetic field, $\lambda_B$, i.e., $\rho_e\gtrsim \lambda_B$. Hereafter, we consider ultra-relativistic particles only, $\gamma\gg1$.

For small deflections, the deflection angle of the velocity (with respect to the particle's initial direction of motion) is approximately the ratio of the change in the electron's transverse momentum to its initial momentum. The former is $ \sim F_L\tau_\lambda$, where $F_L=(e/c)\,{\bf v\times B}$ is the transverse Lorentz force, and $\tau_\lambda$ is the transit time, which is the time required to traverse the the scale of the field's inhomogeneity, i.e., the field correlation length, $\lambda_B$. For an ultra-relativistic particle (whose Lorentz factor is $\gamma \gg 1$), $\tau_\lambda  \sim \lambda_B/c$. The particle's total momentum is, likewise, $p \sim \gamma m_e c$. The change in the transverse momentum is thus, $p_\perp \sim F_L \tau_\lambda \sim e B_\perp \lambda_B/c$. An ultra-relativistic electron will only experience small deviations to its original path; consequently, the deflection angle over the field correlation length will be $\alpha_\lambda \approx p_\perp/p \sim eB_\perp\lambda_B/\gamma m_e c^2$. The subsequent deflection will be in a random direction, because the field is uncorrelated over the scales greater than $\lambda_B$, hence the particle motion is diffusive. As for any diffusive process, the mean squared pitch angle grows linearly with time as
\begin{equation}
\langle \alpha^2 \rangle = D_{\alpha\alpha}t.
\label{diff_def}
\end{equation} 
The pitch-angle diffusion coefficient is, by definition, the ratio of the square of the deflection angle in a coherent patch to the transit time over this patch, that is
\begin{equation}
D_{\alpha\alpha} \sim \frac{\alpha_\lambda^2}{\tau_\lambda}\sim \left(\frac{e^2}{m_e^2 c^3}\right)\frac{\lambda_B}{\gamma^2}{\langle B^2 \rangle},
\label{Daa}
\end{equation}
where an average square magnetic field, $\langle B^2 \rangle$ has been substituted for $B_\perp^2$. Note that the diffusion coefficient depends on both statistical properties of the magnetic field, namely its strength and the typical correlation scale. 

Next, the diffusing particle experiences accelerations and, hence, produces radiation. The radiation from an ultra-relativistic charge is beamed in a cone with a narrow angle of $\Delta{\theta} \sim 1/\gamma$. The ratio of the particle deflection angle to the beaming angle is \citep{medvedev00,medvedev11}:
\begin{equation}
\frac{\alpha}{\Delta\theta} \sim \frac{eB_{\perp}\lambda_B}{m_e c^2} \sim 2\pi \frac{e \langle B^2 \rangle^{1/2}}{m_e c^2 k_B} \equiv \delta.
\label{delta}
\end{equation} 
where we used that $B_{\perp}\sim \langle B^2 \rangle^{1/2}$, as before and the characteristic wavenumber of the turbulence, i.e., $k$ at which $|{\bf B_k}|^2$ has the maximum, is $k_B\sim 2\pi \lambda_B^{-1}$. Note however that the jitter parameter is scale-dependent, therefore at each scale $k$, the jitter parameter is
\begin{equation}	
\delta_k= 2\pi \frac{e \langle B^2_k \rangle^{1/2}}{m_e c^2 k} , 
\label{delta-k}
\end{equation}
where $\langle{B_{k}^2}\rangle^{1/2}$ is the rms field at scale $k$, i.e., it is average over a range of scales from $k$ to $k + \Delta{k}$ with $\Delta{k} \sim k$, that is
\begin{equation}
\langle{B_{k}^2}\rangle = 
{\frac{1}{4\pi}}\int_{{\bf k}}^{{\bf k} +{\bf \Delta{k}}} \!  |{\bf B_{\bf{k}}}|^{2}\mathrm{d}{\bf{k}} \simeq \int_{k}^{2{k}} \! |B_k|^{2}{k^2}\mathrm{d}k.
\label{bk_avg}  
\end{equation} 
The radiation spectrum and diffusion may, generally, be sensitive to this $k$-local value $\delta_k$, though the overall spectrum and diffusion are indeed determined by its averaged value, $\delta$. One important conclusion can readily be made from comparison of Eqs. (\ref{Daa}) and (\ref{delta}). The jitter parameter and the pitch-angle diffusion coefficient are directly related as 
\begin{equation}
D_{\alpha\alpha}\sim\frac{\delta^2}{\gamma^2\tau_\lambda}\sim\frac{\delta^2 k_B c}{2\pi\gamma^2}.
\label{Daa-delta}
\end{equation}

\begin{figure}
\includegraphics[angle = 0, width = 0.9\columnwidth]{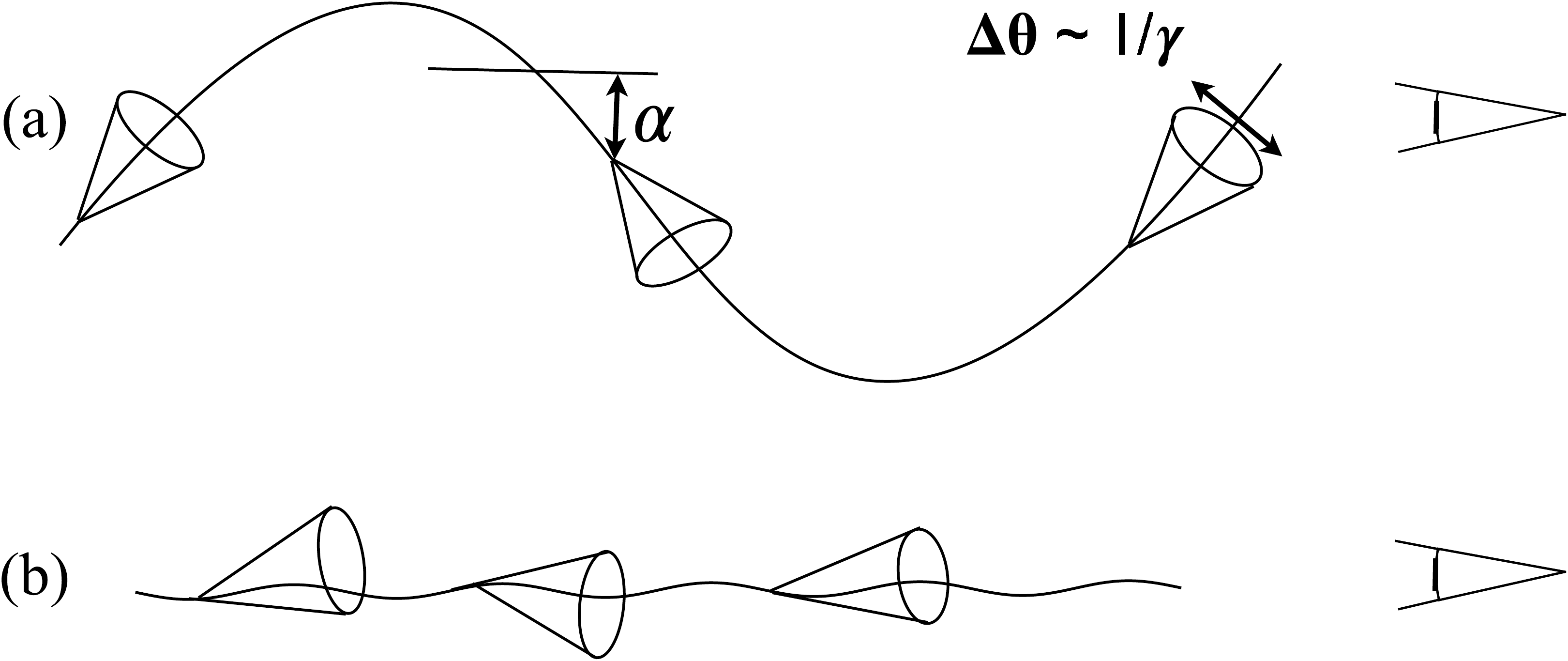}
\vskip-0.1in
\caption{Radiation regimes. (a) Large-angle jitter regime, $1<\delta<\gamma$; radiation is only seen along certain segments of the particle path, thus resulting in the spectrum that is synchotron-like at and above the peak but differing from synchrotron at low frequencies. (b) Small-angle jitter regime, $\delta<1$; radiation is seen from the entire trajectory, thus the spectrum depends on the underlying spectrum of electromagnetic turbulence. }
\label{cartoon}
\end{figure}

There are several regimes of interest. First, $\delta\to\infty$ regime corresponds to the classical synchrotron radiation regime: the particle orbit is a circle in the plane orthogonal to the magnetic field and the field itself is homogeneous. Second, the regime with  $\delta>\gamma$ is very similar to synchrotron, but the particle's guiding center is slowly drifting in an inhomogeneous magnetic field. The produced spectrum is well represented by the synchrotron spectrum in a weakly inhomogeneous magnetic field and it slowly evolves in time due to particle diffusion through regions with different field strength. This regime can be referred to as the diffusive synchrotron regime. Third, when $1<\delta<\gamma$, the particle does not complete it's Larmor orbit because the $B$-field varies on a shorter scale. In this case, an onlooking observer would see radiation from only short intervals of the particle's trajectory (i.e., whenever the trajectory is near the line-of-sight), as in synchrotron, but these intervals are randomly distributed. This is the case of the large-angle jitter regime. The radiation is similar to synchrotron radiation near the spectral peak and above, but differ significantly from it at lower frequencies, see Ref. \citep{medvedev11} for details. Finally, if $\delta<1$, the particle's deviations are extremely small compared to its beaming angle, so the radiation seen by an observer is produced over a large number of incoherent patches of the $B$-field, i.e., almost throughout the entire trajectory of the particle \citep{medvedev00, medvedev06, medvedev11}, as is shown in Fig. \ref{cartoon} . Thus, the resulting radiation markedly differs from synchrotron radiation, although the total radiated power of radiation, $P_\text{tot}\equiv dW/dt$, produced by a particle in all these regimes, e.g., jitter and synchrotron, are identical: 
\begin{equation}
P_\text{tot} = \frac{2}{3} r_e^2c \gamma^2 \langle B^2 \rangle\sim\frac{e^2}{c}\, \frac{\omega_c^2}{\gamma^2}, 
\end{equation}
where $r_e = e^2/m_e c^2$ is the classical electron radius and $\omega_c\sim\gamma^2 e\langle B^2\rangle^{1/2}/m_e c$ is the characteristic synchrotron frequency of radiation in a uniform or weakly inhomogeneous (on the Larmor scale) magnetic field.

Radiation produced in the small-angle jitter regime, $\delta<1$, is the most sensitive to and provides the most information on the turbulence properties, hence, we mostly consider this regime in our paper. The characteristic frequency of radiation can be estimated by considering the virtual photon approximation. The correlation scale $\lambda_{B}'$, as seen in the electron's rest frame, is related to $\lambda_B$ in the lab frame by the Lorentz transformation (i.e., $\lambda_{B}' = \lambda_{B}/\gamma$). The electron's transverse acceleration occurs over the length scale of $\lambda_B$, which corresponds to a characteristic time scale, in the electron's frame, of $\tau' \sim \lambda_B'/c$. Thus, the radiation emitted by the electron has a characteristic frequency, in the electron's rest frame, of $\omega' \sim 2\pi/\tau' \sim 2\pi{c}/\lambda_B' \sim 2\pi\gamma{c}/\lambda_{B}$. Transforming back to the lab frame picks up another factor of $\gamma$ from the Lorentz transformation, giving the characteristic frequency of $ \omega_j \sim 2\pi{c}\gamma^2/\lambda_B$, or:
\begin{equation}
\omega_j  \sim \gamma^2 k_B c \sim 2\pi \omega_c/ \delta .
\label{omegaj}
\end{equation} 
The spectral power at the characteristic (spectral break) frequency is
\begin{equation}
P(\omega_j)\sim \frac{P_\text{tot}}{\omega_j}\sim\frac{e^2}{4\pi^2 c}\frac{\omega_j\delta^2}{\gamma^2}.
\label{Pomegaj}
\end{equation}
Note that this equation allows one to determine the jitter parameter directly from spectral observations:
\begin{equation}
\delta\sim \frac{2\pi c^{1/2}}{e}\left[\frac{\gamma^2P(\omega_j)}{\omega_j}\right]^{1/2}.
\label{delta-Pomegaj}
\end{equation}

From Eqs. (\ref{Daa-delta}), (\ref{omegaj}) one obtains the relation between the spectral break of radiation and the pitch-angle diffusion coefficient:
\begin{equation}
D_{\alpha\alpha}\sim\frac{\omega_j\delta^2}{2\pi\gamma^4}.
\label{Daa-omegaj}
\end{equation}
One can also relate the diffusion coefficient to the spectral power at the break frequency from Eq. (\ref{Pomegaj}). We obtain
\begin{equation}
D_{\alpha\alpha}\sim\frac{2\pi c}{e^2}\frac{P(\omega_j)}{\gamma^2}.
\label{Daa-spectrum}
\end{equation}
We emphasize that $D_{\alpha\alpha}$ is, thus, a directly measurable quantity because it solely depends on the radiation spectral power at the break frequency, $P(\omega_j)$, and the Lorenz factor, $\gamma$, of the radiation-emitting electrons. Thus, radiative diagnostics should be a very useful technique to study laboratory plasmas and explore plasma conditions in astrophysical plasmas.

In general, turbulence has a range of scales rather than a single scale $k_B$. We explore how spectral and transport properties are modified in this more realistic case. In the analysis, we assume the isotropic three-dimensional magnetic turbulence with a power law turbulent spectrum:
\begin{equation}
|\textbf{B}_\textbf{k}|^2 \propto k^{-\mu},
\label{Bk}
\end{equation} 
if $k_\text{min}\le k\le k_\text{max}$ and zero otherwise. Here $\textbf{B}_\textbf{k}$ is the spatial Fourier transform of the magnetic field, $\textbf{k}$ is the wave vector, and index $\mu$ is a positive number. Notice that we consider a static isotropic turbulence, hence the field distribution is independent of time and the wave vector direction. Note that such magnetic turbulence is a natural outcome of the non-linear Weibel/filamentation instability, which occurs at relativistic collisionless shocks and in laser-produced plasmas. It has been shown \citep{medvedev06, medvedev11,RK10,TT11} that monoenergetic relativistic electrons in such turbulence produce flat angle-averaged spectra
below the spectral (jitter) break and power-law spectra above the break, that is
\begin{equation}
P(\omega) \propto 
\left\{\begin{array}{ll}
\omega^0, &\text{if}~ \omega<\omega_j, \\
\omega^{-\mu + 2}, &\text{if}~ \omega_j<\omega<\omega_b, \\
0, &\text{if}~ \omega_b<\omega,
\end{array}\right.
\label{Pomega}
\end{equation}
where the spectral (jitter) break is defined in this case as 
\begin{equation}
\omega_j =\gamma^2 k_\textrm{min} c, 
\label{omegaj-kmin}
\end{equation}
because the field $B_k$ at $k_\text{min}$ has the most power (for $\mu>0$) and, hence, $k_\text{min}$ plays the role of $k_B$ in Eq. (\ref{omegaj}). Similarly, the high-frequency break is 
\begin{equation}
\omega_b =\gamma^2 k_\textrm{max} c.
\label{omegab}
\end{equation}
Thus, by observing the radiation spectrum, one can probe the structure of the magnetic turbulence that generates it. Note that the angle-averaged single electron spectra are equivalent to the ensemble-averaged spectra for isotropic particle distribution functions. We will use the latter in our numerical study.

\section{Numerical model}
\label{s:model}

Our goal is to explore, via simulations of particle dynamics in magnetic turbulence, the diffusive and radiative properties of plasmas and how they are related. We perform our simulations from first-principles. We use a 3D simulation box with periodic boundary conditions in all directions. Relativistic electrons are test particles moving in the preset magnetic fields but do not interact with each other, nor do they induce any fields. Radiative energy losses are considered negligible compared to the energies of individual particles, hence neglected. Motion of each electron is, thus, solely determined by the Lorentz force equation given by:
\begin{equation}
\frac{d{\boldsymbol\beta}}{dt} = -\frac{e}{\gamma m_ec}\ {\boldsymbol\beta}\times\textbf{B},
\label{dvdt}
\end{equation}
where  ${\boldsymbol\beta} \equiv {\bf v}(t)/c$.
Note that this Lorentz force conserves particle's energy, therefore the magnitude of the velocity is constant in time and only the direction is changing. The simulation can be divided into two principle stages (see Ref \citep{keenan12} for a detailed description of the numerical implementation). First, the turbulent magnetic field is generated from a given spectral distribution in Fourier space. This field is created on a lattice that is then interpolated, so that a ``continuous" field is represented. The interpolation was implemented by way of divergenceless matrix-valued radial basis functions (see Ref. \citep{mcnally11}, for a discussion). This method begins with a radial function -- in our case, one of the simplest, $\phi({\bf r}) = e^{-\epsilon{r}}$ (where $\epsilon$ is a scaling factor, and $r^2 = x^2 + y^2 + z^2$). Then, a set of divergence-free matrix-valued radial basis functions is obtained from the transformation \citep{mcnally11}:
\begin{equation}
\Phi({\bf r}) = (\nabla\nabla^T - \mathbb{I}_{3\times3}{\nabla^2})\phi({\bf r}),
\label{rad_basis}
\end{equation} 
where $\nabla\nabla^T$ is the second-order, $3\times3$-matrix differential operator and $\mathbb{I}_{3\times3}$ is the $3\times3$ identity matrix. These interpolants were applied to the interior of each lattice ``cube" (i.e.\ the space between a single lattice point and the five immediately adjacent points).  The second stage in our model then involves the numerical solution of the equation of motion for each particle, from which $\langle\alpha^2\rangle$ and the radiation spectra are obtained. We will first turn our attention to the generation of the magnetic field.

Generation of the magnetic field distribution is more convenient in Fourier space for two reasons. Firstly, it is simpler to specify a particular spectral distribution in Fourier space directly, rather than attempting to find a field in real space that would fit the chosen spectral distribution. Secondly, any physically realizable field should satisfy Maxwell's equations, thus its divergence must be zero. Producing a divergenceless field in Fourier space is far simpler than generating one in real space. This is because Gauss' law, $\nabla\cdot{\bf B}=0$, is an algebraic equation in Fourier space, ${\bf k}\cdot{\bf B_k}=0$, for each Fourier component, ${\bf B_k}$, corresponding to the wave vector ${\bf k}$. Although our code can handle a wide variety of magnetic spectral distributions, this study is limited to the case of isotropic magnetic turbulence, described in Eq. (\ref{Bk}), and we have left consideration of more sophisticated models for the future.

The next stage in the simulation starts with the numerical solution of the equation of motion, Eq. (\ref{dvdt}). This was done via a fixed step 4$^\text{th}$-order Runge-Kutta-Nystr\"om method. With all the particle positions, velocities, and accelerations calculated, the radiation spectrum is obtained from the equation \citep{landau75,jackson99}
\begin{equation}
\frac{d^2W}{d\omega\, d\Omega} = 
\frac{e^2}{4\pi c^2}  \left|\int_{-\infty}^\infty \! \frac{\hat{\bf n}\times[(\hat{\bf n} - {\boldsymbol\beta}) \times \dot{\boldsymbol\beta} ]}{(1 - \hat{\bf n}\cdot{\boldsymbol\beta})^2} e^{i\omega\left(t - \hat{\bf n}\cdot {\bf r}(t)/c \right)}\, \mathrm{d} t
\right|^2,
\label{LW}  
\end{equation} 
which represents the radiative spectral energy, $dW$ per unit frequency, $d\omega$, and per unit solid angle, $d\Omega$. In this equation $\textbf{r}(t)$ is the particle's position at the retarded time $t$, $\hat{\bf n}$ is the unit vector pointing from  $\textbf{r}(t)$ to the observer and $\dot{\boldsymbol\beta} \equiv \text{d}{\boldsymbol\beta}/\text{d}t$. There is some freedom in the calculation of the total radiation spectrum. One can add the spectra coherently (i.e., by taking the Fourier transform of the ``summed over" radiation fields of each particle). In this case, only a single integration would be needed. Alternatively, we can add the spectra incoherently (i.e., by integrating each particle's radiation field separately, and then summing the results of each integration). Both approaches result in the same spectra, but those obtained in the former approach are noisier for the same number of simulation particles. Hence we use the second approach in our study. 

\begin{figure}
\includegraphics[angle = 0, width = 1\columnwidth]{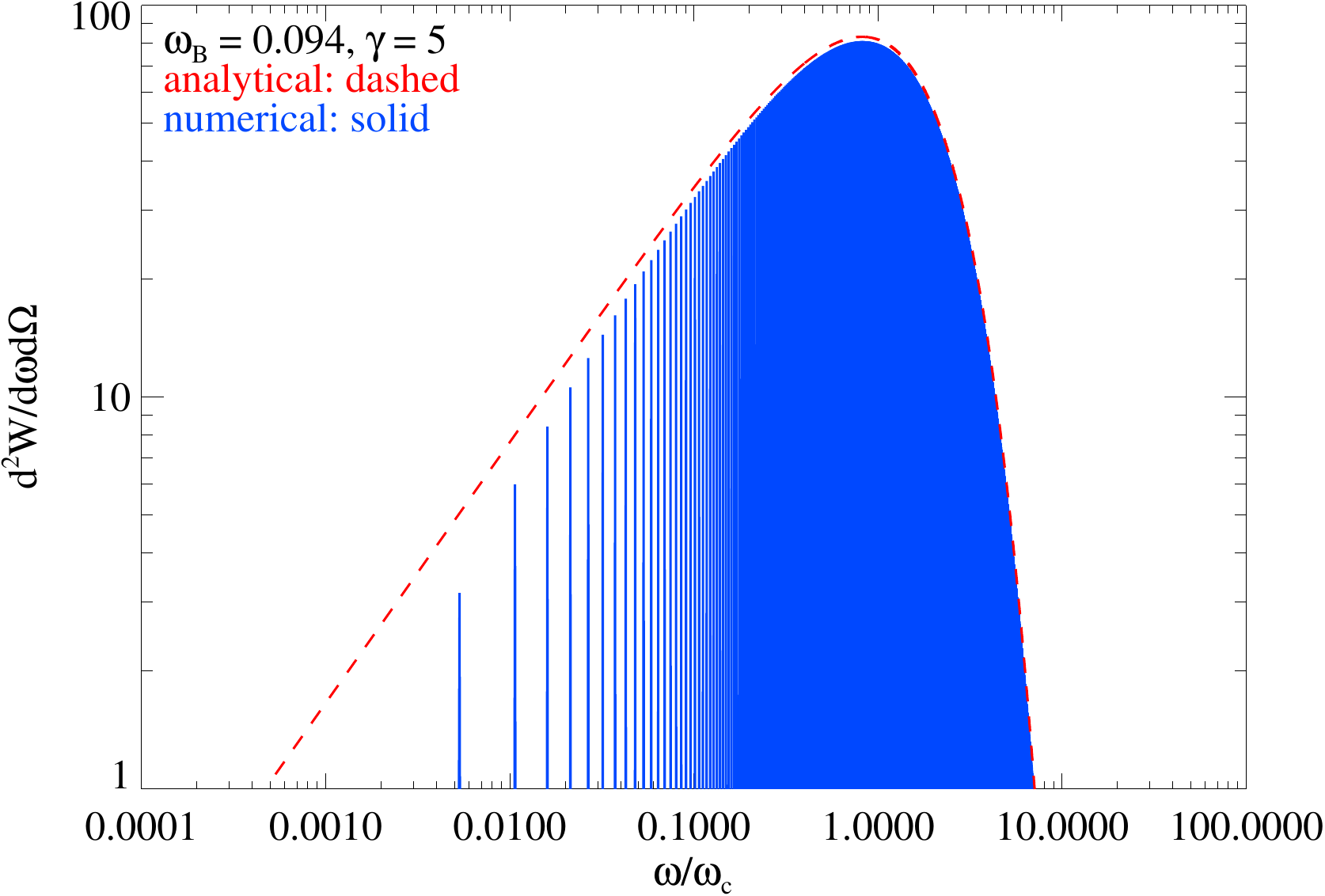}
%\vskip-0.1cm
\caption{Radiation spectrum (i.e., the total radiated energy, $dW$ per unit solid angle, $d\Omega$ per unit frequency, $d\omega$ vs. frequency, $\omega$) of a single relativistic charge moving through a uniform magnetic field. The numerical solution is indicated in blue, the red line is the analytical solution, and they agree very well. The spectrum is peaked at the synchrotron frequency  $\omega_c = (3/2)\gamma^2\omega_B$, where $\omega_B = eB_{\perp}/\gamma{m_e}{c}$ is the electron gyrofrequency. In this and other spectral plots, the radiation power is arbitrarily normalized. }
\label{synch}
\end{figure}

Our code was tested in various set-ups. In one, we considered motion of a single relativistic particle in a uniform magnetic field. Figure \ref{synch} shows the obtained radiation spectrum and the corresponding theoretical synchrotron spectrum. Here, the blue curve represents the numerically resolved synchrotron harmonics (which are integer multiples of the gyrofrequency, $\omega_B = eB_{\perp}/{m_e}c$). We see excellent agreement with the analytical result, indicated in red. 

\begin{figure}
\includegraphics[angle = 0, width = 1\columnwidth]{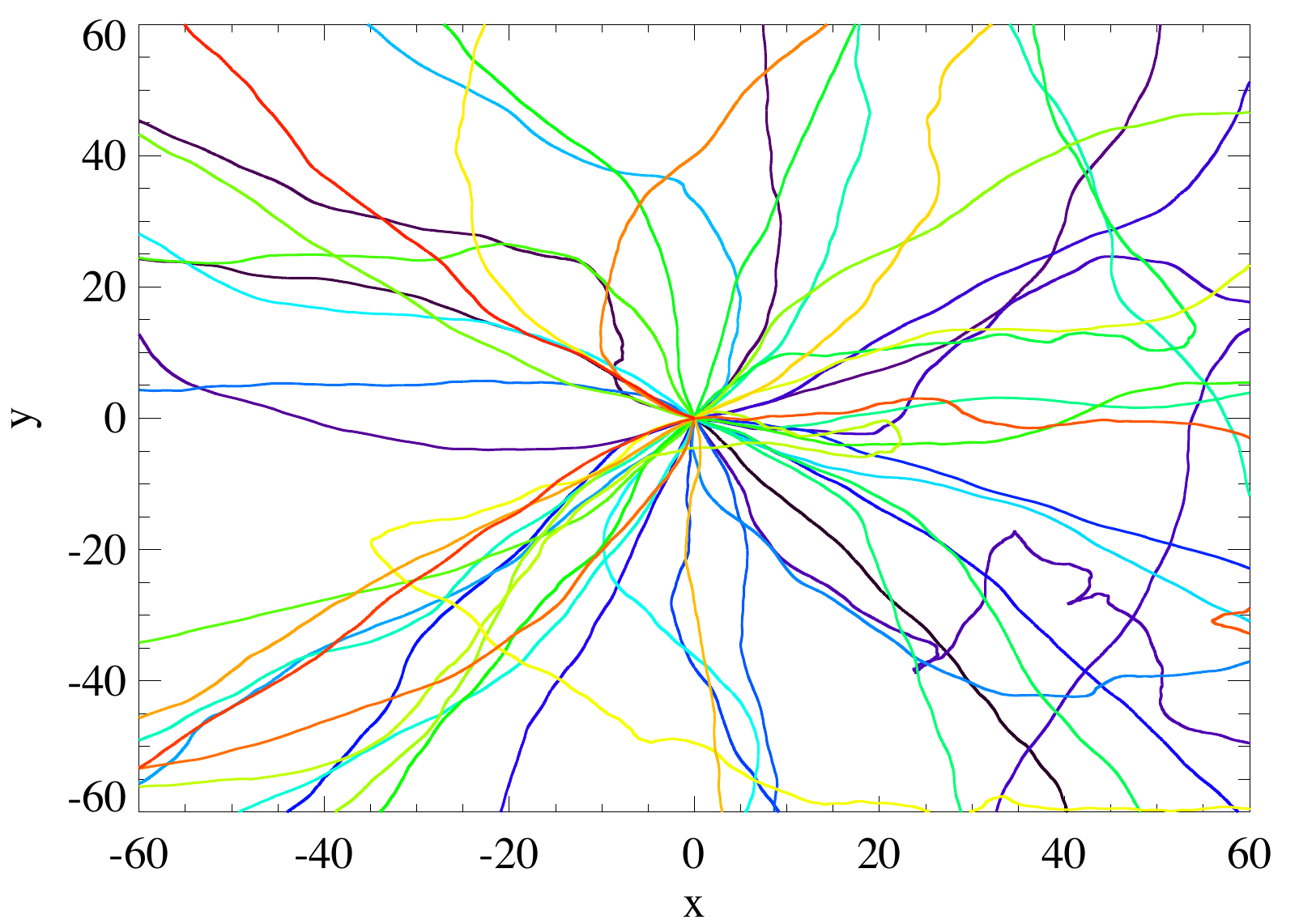}
%\vskip-0.1cm
\caption{The trajectories of 50 monoenergetic ($\gamma = 3$) particles through a turbulent magnetic field ($\delta \sim 1$) projected on to the $x$-$y$ plane.  Each particle (denoted by a unique color) starts from an origin with a random initial velocity. The axes are $x$ and $y$ positions in simulation units.}
\label{traj}
\end{figure}
\begin{figure}
\includegraphics[angle = 0, width = 1\columnwidth]{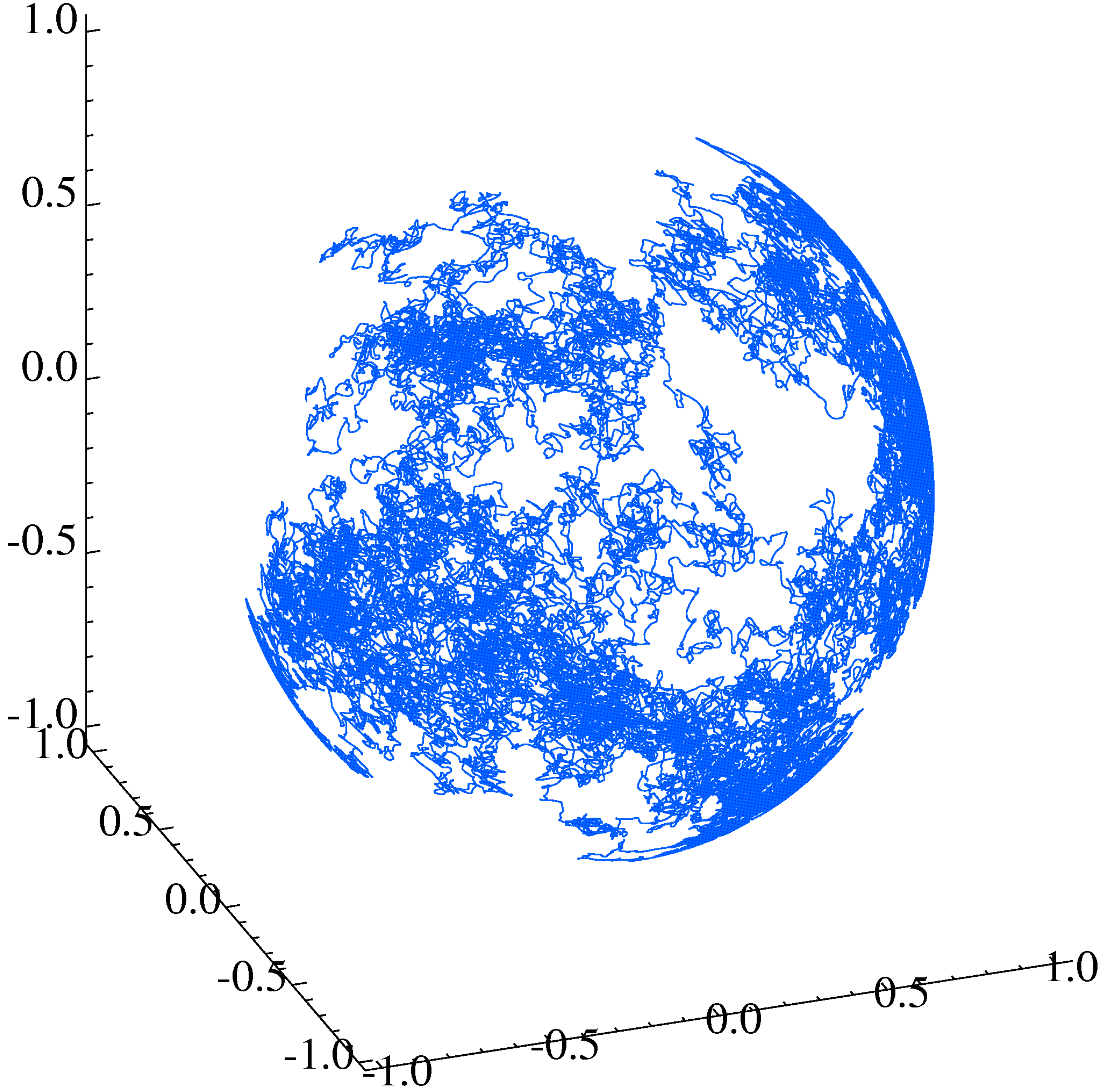}
%\vskip-0.1cm
\caption{Velocity space of a single particle ($\gamma = 5$) moving through isotropic magnetic turbulence ($\delta \sim 1$). The axes are the components of the velocity, which are in units of $c$. Notice that, although the velocity vector of the particle varies randomly (and, over enough time, visits all possible directions), its magnitude is constant. }
\label{v_space}
\end{figure}
\begin{figure}
\includegraphics[angle = 0, width = 1\columnwidth]{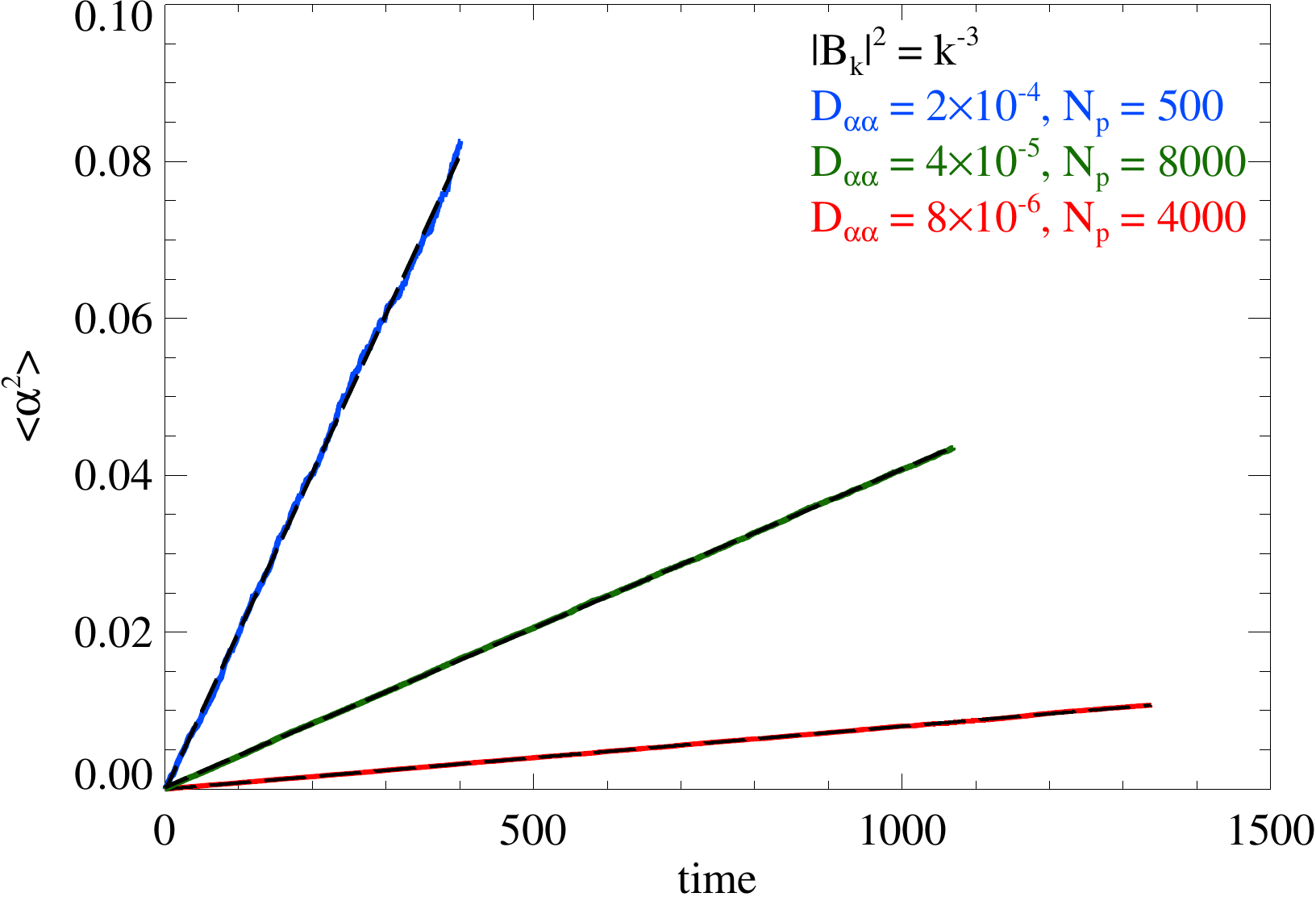}
%\vskip-0.1cm
\caption{Average square pitch-angle vs. time (in simulation units). Here, $D_{\alpha\alpha}$ is the numerically obtained pitch-angle diffusion coefficient (i.e., the slope of the line). The slopes are estimated via lines of best fit (indicated by black dashed lines). Notice that, although these results differ by orders of magnitude in the slope (and, additionally, particle number), there is not an appreciable deviation from linear behavior (as long as $\delta \ll \gamma$).}
\label{diff_slope}
\end{figure}

We also verified that motion of a particle in small-scale random magnetic fields results in pitch-angle diffusion and the particle orbit is chaotic. The chaotic nature inherent in the particle motion is illustrated in Figure \ref{traj}. Here, 50 monoenergetic ($\gamma = 3$) particles are sent out from an origin in random directions. The variability in the particle motion is seen after shortly leaving the origin. The diffusive nature of the particle motion is evident from Figure \ref{v_space} and Figure \ref{diff_slope}. In Figure \ref{v_space}, the velocity space of a single relativistic particle ($\gamma = 5$) is plotted over a very large simulation time. Notice that the particle velocity is confined to a sphere (i.e.,\ $\gamma$ is constant), and that the velocity vector diffusively visits various directions in the course of the particle's motion. In Figure \ref{diff_slope}, the average square pitch-angle, $\langle\alpha^2\rangle$ is plotted versus $t$ for three different cases. The linear dependence is indicative of pitch-angle diffusion, cf. Eq. (\ref{diff_def}). In all three cases, $\delta \approx 1$. Note that the diffusion approximation of these particles' motion is accurate so long as $\delta \ll \gamma$.

Finally, we verified the numerical convergence of the results by changing the size of the simulation box and the total number of simulation particles. Because of vastly different computational requirements in various cases, however, we were sometimes forced to use a smaller number of particles than in other scientific runs. This introduced more noise in some spectra presented below.  

\section{Numerical results}

\label{s:results}

In Section \ref{s:analytic} we made a number of theoretical predictions concerning radiation and transport properties of plasmas with small-scale turbulent magnetic fields. Here we check these results with numerical simulations and then further explore how radiation spectra depend on underlying magnetic field distributions. 

First of all, we explore how the pitch-angle diffusion coefficient depends on various parameters, cf. Eq. (\ref{Daa}), namely the particle's Lorentz factor, $\gamma$, the magnetic field strength $\langle B^2 \rangle$ and the jitter parameter $\delta$. 

\begin{figure}
\includegraphics[angle = 0, width = 1\columnwidth]{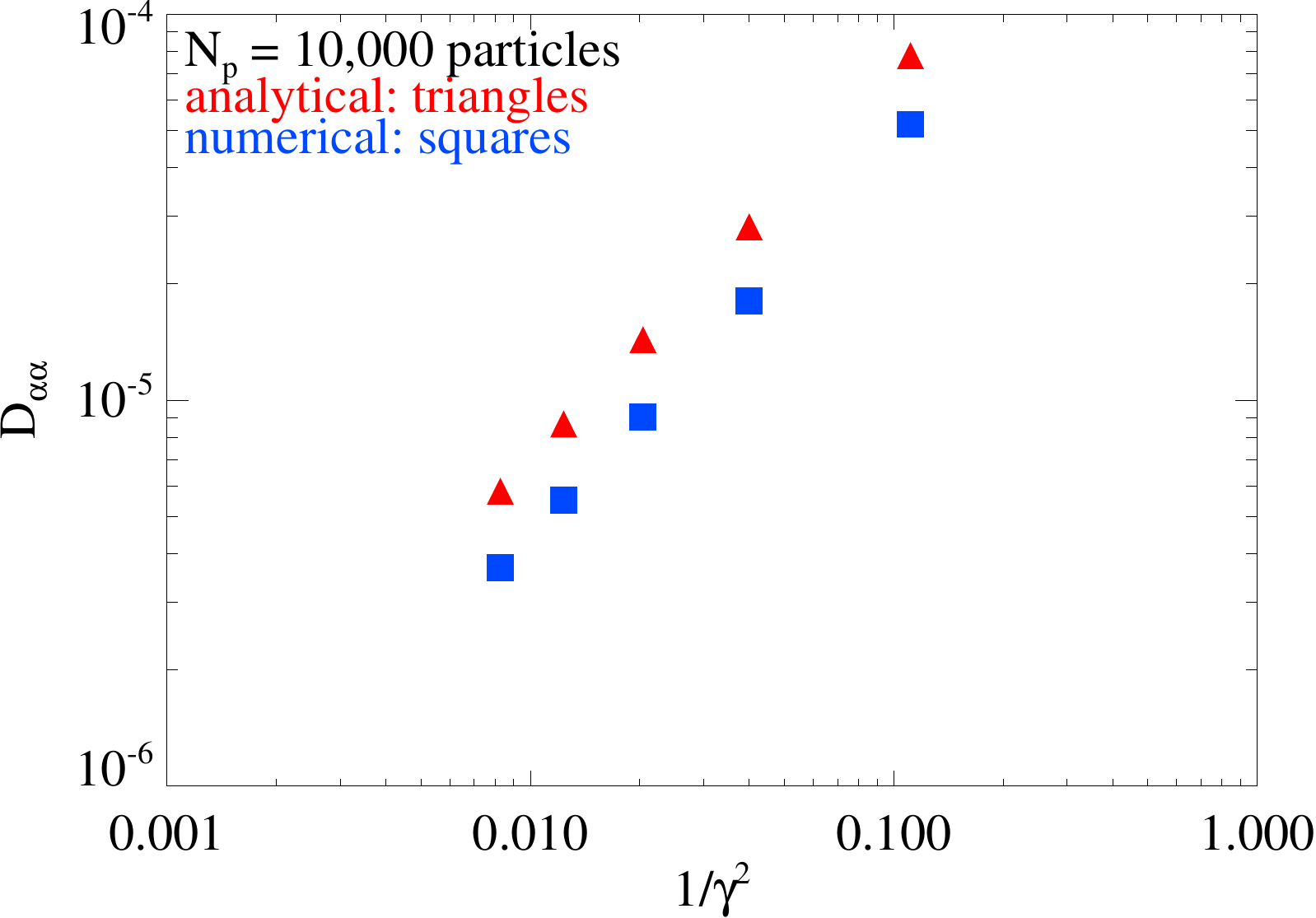}
%\vskip-0.1cm
\caption{Pitch-angle diffusion coefficient, $D_{\alpha\alpha}$ vs. particle inverse-square Lorentz factor, $1/\gamma^{2}$. The ``blue squares'' indicate the $D_{\alpha\alpha}$ obtained directly from simulation (as the slope of $\langle\alpha^2\rangle$ vs. time), while the ``red triangles" are the analytical, given by Eq. (\ref{Daa-omegaj}), pitch-angle diffusion coefficients. }
\label{diff_gamma}
\end{figure}

\begin{figure}
\includegraphics[angle = 0, width = 1\columnwidth]{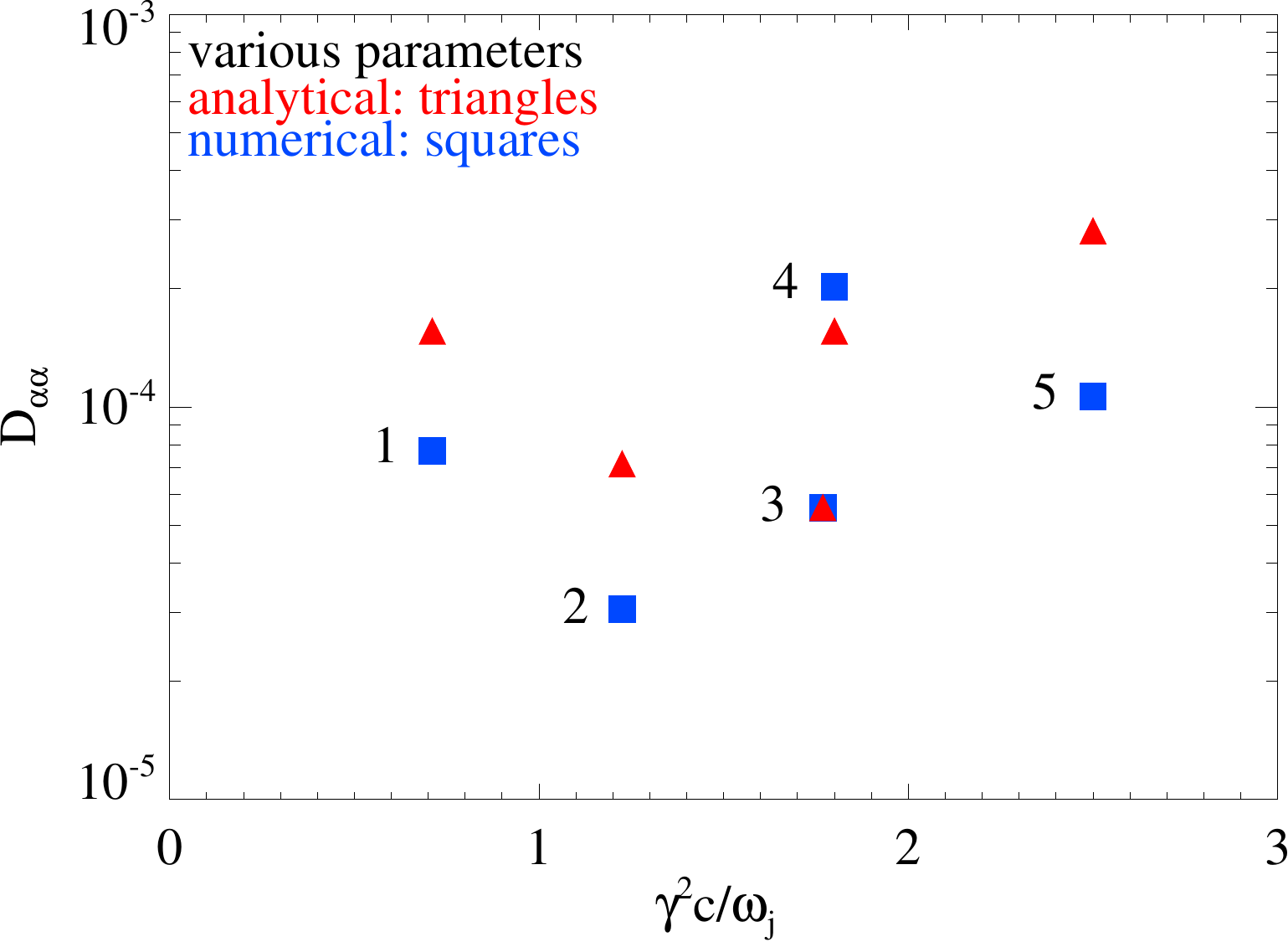}
%\vskip-0.1cm
\caption{$D_{\alpha\alpha}$ vs. the frequency of $\omega_j$, i.e., the frequency at which the radiation spectrum peaks: Eq. (\ref{omegaj}) (in this case, obtained numerically). Once more, the ``blue squares" indicate the $D_{\alpha\alpha}$ obtained directly from simulation while the ``red triangles" are the analytical $D_{\alpha\alpha}$, given by Eq. (\ref{Daa-omegaj}).  This figure represents cases with various values of $k_\text{min}$, $k_\text{max}$, $N_p$, $\langle B^2 \rangle$, $\Delta{t}$, and $\gamma$, hence the large spread in $D_{\alpha{\alpha}} $'s (see Table \ref{parameter_table} for a listing of the parameters). Notice that, although there is considerable variation between the difference in the analytical and numerical diffusion coefficients, the difference is never greater than a factor of a few.}
\label{diff_peak}
\end{figure}

\begin{figure}
\includegraphics[angle = 0, width = 1\columnwidth]{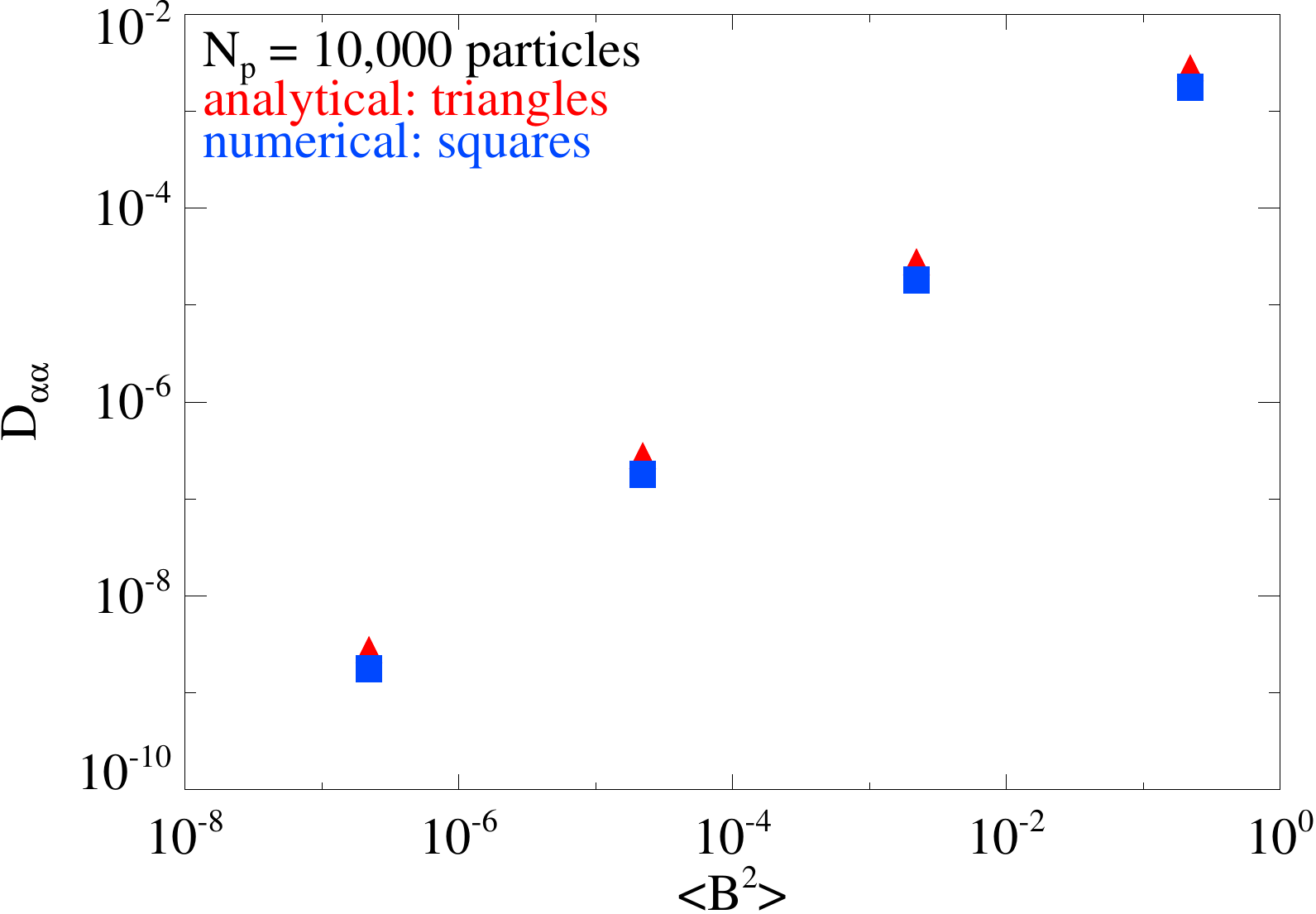}
%\vskip-0.1cm
\caption{$D_{\alpha\alpha}$ vs. $\langle B^2 \rangle$. As before, the ``blue squares" indicate the $D_{\alpha\alpha}$ obtained directly from simulation while the ``red triangles" are the analytical $D_{\alpha\alpha}$, given by Eq. (\ref{Daa-omegaj}). }
\label{diff_b}
\end{figure}

Figure \ref{diff_gamma} shows the $\gamma$ dependence of $D_{\alpha\alpha}$. Close agreement between the theoretical and numerical diffusion coefficients with varying Lorentz factor is evident. Next, $D_{\alpha\alpha}$ depends on two additional parameters, namely $k_\text{min}$ and $\langle B^2 \rangle$. Instead of plotting $D_{\alpha\alpha}$ vs. $k_\text{min}$ directly, we chose to plot the diffusion coefficient vs. the frequency (Figure \ref{diff_peak}) of $\omega_j$ -- which is given in Eq. (\ref{omegaj}) -- in the radiation spectrum. 

Figure \ref{diff_peak} represents cases with different values of  $k_\text{min}$'s, $k_\text{max}$'s, $\Delta{t}$'s, $N_\text{p}$'s, $\langle B^2 \rangle$'s,  and $\gamma$'s, which is why the points show large spread (the parameters used for each data point are listed in Table \ref{parameter_table}). The dependence of $D_{\alpha\alpha}$ on the magnetic field strength, $\langle B^2 \rangle$, is illustrated in Figure \ref{diff_b}, which agrees with the prediction given by Eq. (\ref{Daa}). It is worth noting that the apparent difference between the theoretical and numerical diffusion coefficients in these plots is due to the approximate nature of our theoretical analysis in Section \ref{s:analytic}.

We now present radiation spectra and demonstrate how they are related to the parameters of the turbulent magnetic field. In all runs, we specify the field strength with the jitter parameter, given by Eq. (\ref{delta}). Also, we will plot the angle-integrated spectra $dW/d\omega$ rather than $dW/d\Omega{d\omega}$. The former represent the spectra from an ensemble of particles with an isotropic velocity distribution. Thus, by summing the individual spectra of each particle, the solid angle, $d\Omega$ has been effectively ``integrated'' out.

% Requires the booktabs if the memoir class is not being used
\begin{table}
   \centering
   %\topcaption{Table captions are better up top} % requires the topcapt package
   \begin{tabular}{l*{9}{c}r} % Column formatting, @{} suppresses leading/trailing space
      \toprule
      %\cmidrule(r){1-2} % Partial rule. (r) trims the line a little bit on the right; (l) & (lr) also possible
    \midrule
    \hline
     $\#$ & $\delta$ & $\Delta{t}$ & $\gamma$ & $\mu$ & $k_\text{min}$ & $k_\text{max}$ & $\langle B^2 \rangle^{1/2}$ & $N_p$ \\
     \hline
     \midrule
$1$ & $0.12$ & $0.0025$ & $5$ & $-3$ &  $1.3$ & $32.2$ & $0.024$ &  $4000$    \\
$2$ & $0.63$ & $0.0100$ & $8$ & $-3$ &  $1.0$ & $16.1$ & $0.100$ & $2000$ \\
$3$ & $0.47$ & $0.0100$ & $7$ & $-3$ &  $0.6$ & $16.1$ & $0.047$ & $500$ \\
$4$ & $0.47$ & $0.0100$ & $3$ & $-3$ &  $0.6$ & $16.1$ &  $0.047$ & $500$  \\
$5$ & $0.94$ & $0.0100$ & $5$ & $-3$ &  $0.3$ & $16.1$ & $0.047$ & $500$ \\
      \bottomrule
   \end{tabular}
\caption{Table of parameters used in Figure \ref{diff_peak}.}
\label{parameter_table}
\end{table}

\begin{figure}
\includegraphics[angle = 0, width = 1\columnwidth]{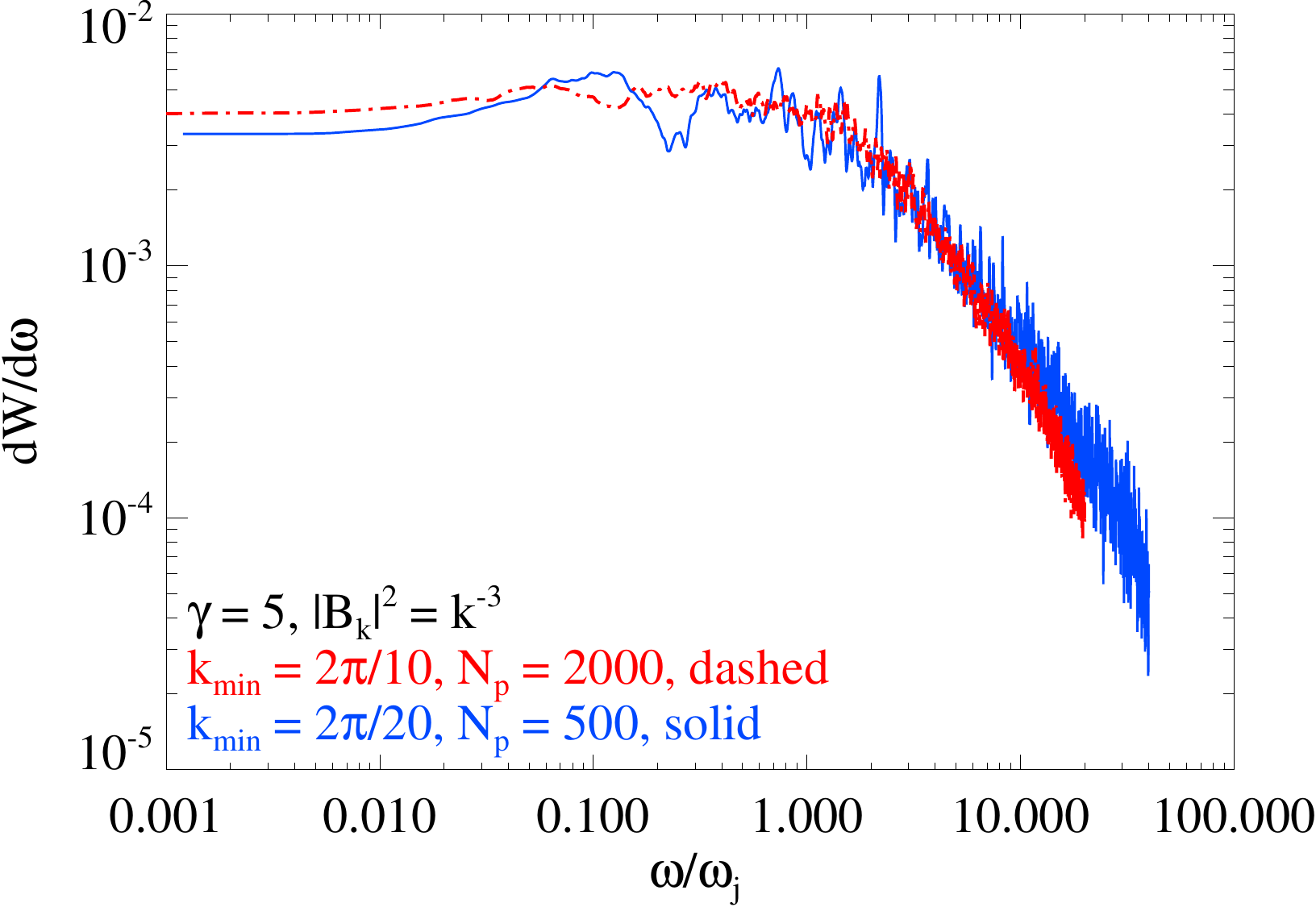}
%\vskip-0.2cm
\caption{Radiation spectrum ($dW/d\omega$ vs $\omega$) illustrating the $k_\text{min}$ dependence. Recall that $k_\text{min}$ is the characteristic wave number of the turbulent magnetic field (given by a negatively sloped power law spectral distribution). The scaling with $k_\text{min}$ agrees with Eq. (\ref{omegaj-kmin}). }
\label{kmin_comp}
\end{figure}

\begin{figure}
\includegraphics[angle = 0, width = 1\columnwidth]{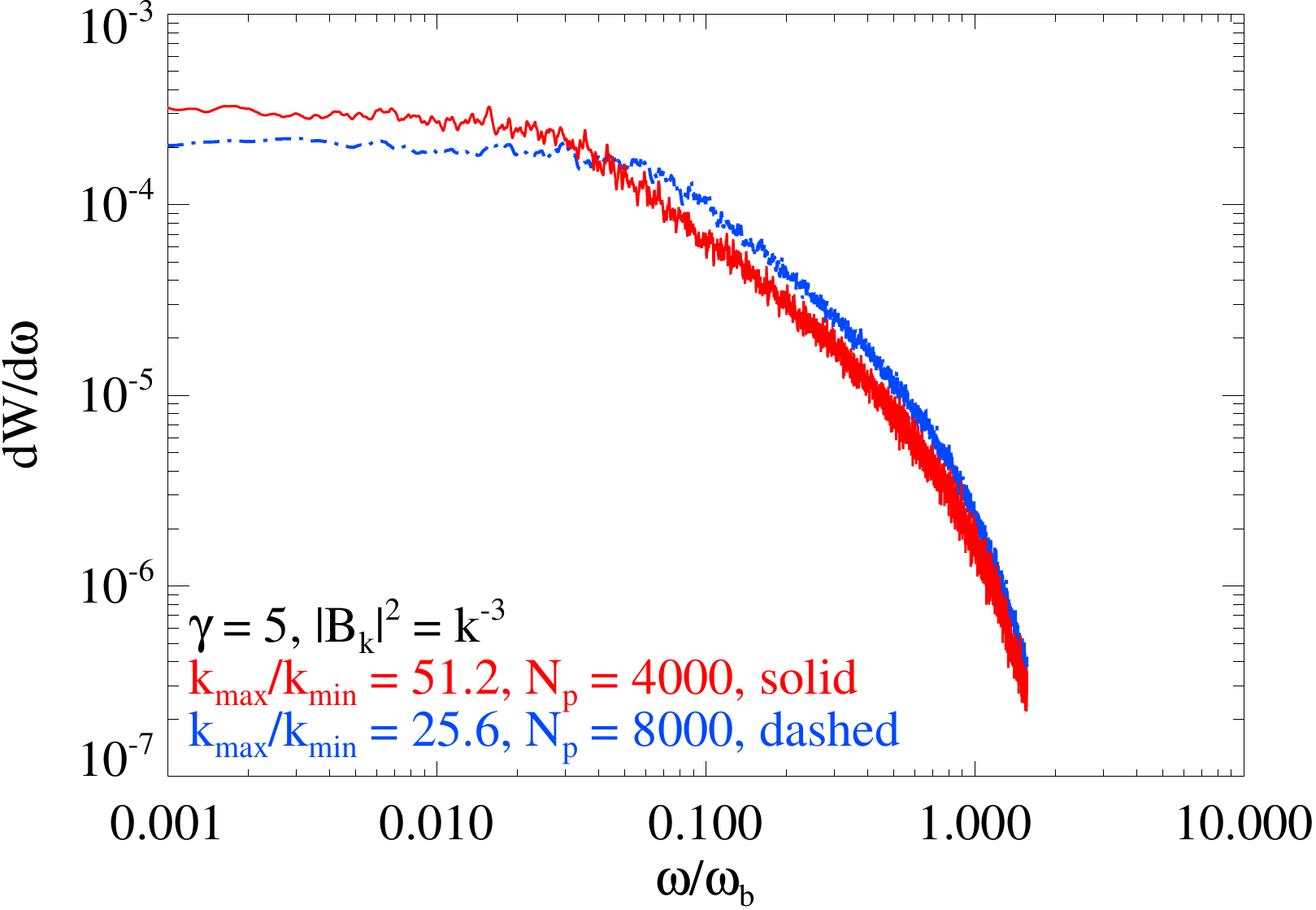}
%\vskip-0.2cm
\caption{Radiation spectrum illustrating the $k_\text{max}$ dependence (the two spectra differ in $k_\text{max}$ by a factor of 2). The transition from a power law to a steep drop off occurs at $\omega_b \sim \gamma^2k_\text{max}c$. }
\label{kmax_comp}
\end{figure}

\begin{figure}
\includegraphics[angle = 0, width = 1\columnwidth]{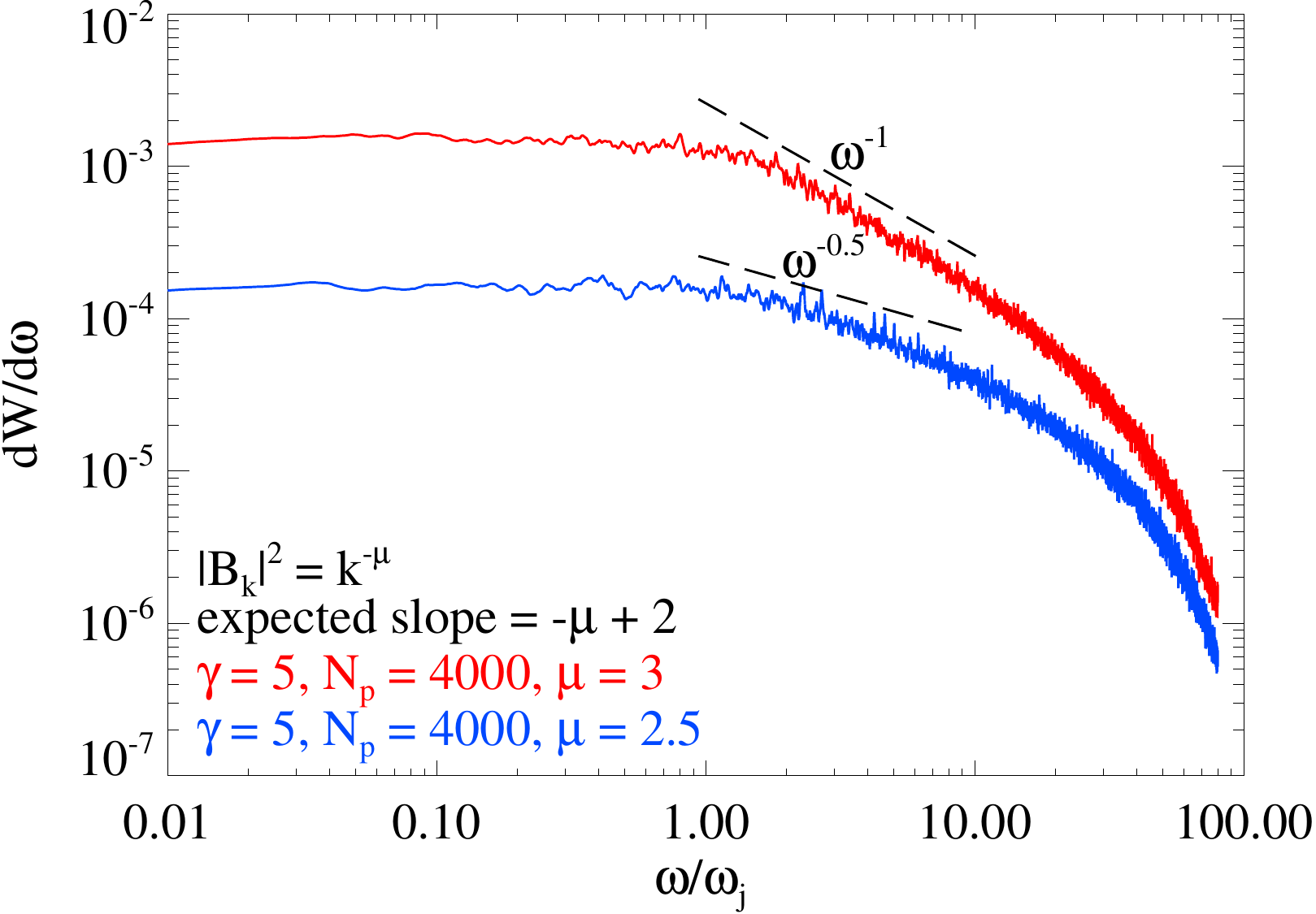}
%\vskip-0.2cm
\caption{An illustration of the magnetic spectral index, $\mu$ dependence. Given a magnetic spectral distribution $|B_k|^2 = k^{-\mu}$, where $\mu$ is positive, the expected slope (in a log-log plot) of the particle radiation spectrum is $-\mu + 2$; following $\omega_j = \gamma^2k_\text{min}c$, it will be zero for lesser frequencies, in agreement with Eq. \eqref{Pomega}.  }
\label{mu_comp}
\end{figure}

Figure \ref{kmin_comp}, shows the radiation spectra as a function of the normalized frequency, $\omega/\omega_j$ for two different values of $k_\text{min}$. Other parameters are: $\gamma = 5$, $\mu = 3$, $\langle B^2 \rangle$ $= 0.047$, and $k_\text{max} = 128\pi/25$ (in both runs), $\delta = 0.47$, $N_p = 500$ (in the $k_\text{min} = \pi/5$ case), $\delta = 0.94$, and $N_p = 2000$ (in the $k_\text{min} = \pi/10$ case). The spectral break occurs somewhat below $\omega_j$ but it correctly scales with $k_\text{min}$, cf Eq. (\ref{omegaj-kmin}). The spectrum is flat below the spectral break and falls-off above it roughly as $\omega^{-1}$, which is expected for the used value of $\mu=3$. There is a hint of the second break at the high-frequency end due to the high-$k$ cut-off of the magnetic spectrum. To demonstrate this we performed two long runs with different values of $k_\text{max}/k_\text{min}=25.6,\ 51.2$ (other parameters being the same). Figure \ref{kmax_comp} shows the results of these runs. One can see that the extent of the power-law part of spectrum increases with increasing value of $k_\text{max}/k_\text{min}$. The frequency in the figure is normalized by $k_\text{max}$ rather than $k_\text{min}$ and the alignment of the high-frequency breaks is evident. Thus, this break is indeed due to the small-scale cut-off of the magnetic field spectrum, cf Eq. (\ref{omegab}). The spectral slope in the power-law region between the two breaks is determined by the magnetic field spectral index $\mu$. Figure \ref{mu_comp} shows the results of two runs with two values of $\mu=2.5,\ 3$. Other parameters, for both cases, are: $\gamma = 5$, $k_{min} = \pi/5$, $k_{max} = 256\pi/25$, $N_p = 4000$, and $\langle B^2 \rangle$ $= 0.047$. One sees that the spectral power law index is indeed close to the value of $-\mu + 2$, as is given by Eq. (\ref{Pomega}). Thus, the radiation spectrum in the small-angle jitter regime is largely determined by the $B$-field spectrum.

\begin{figure}
\includegraphics[angle = 0, width = 1\columnwidth]{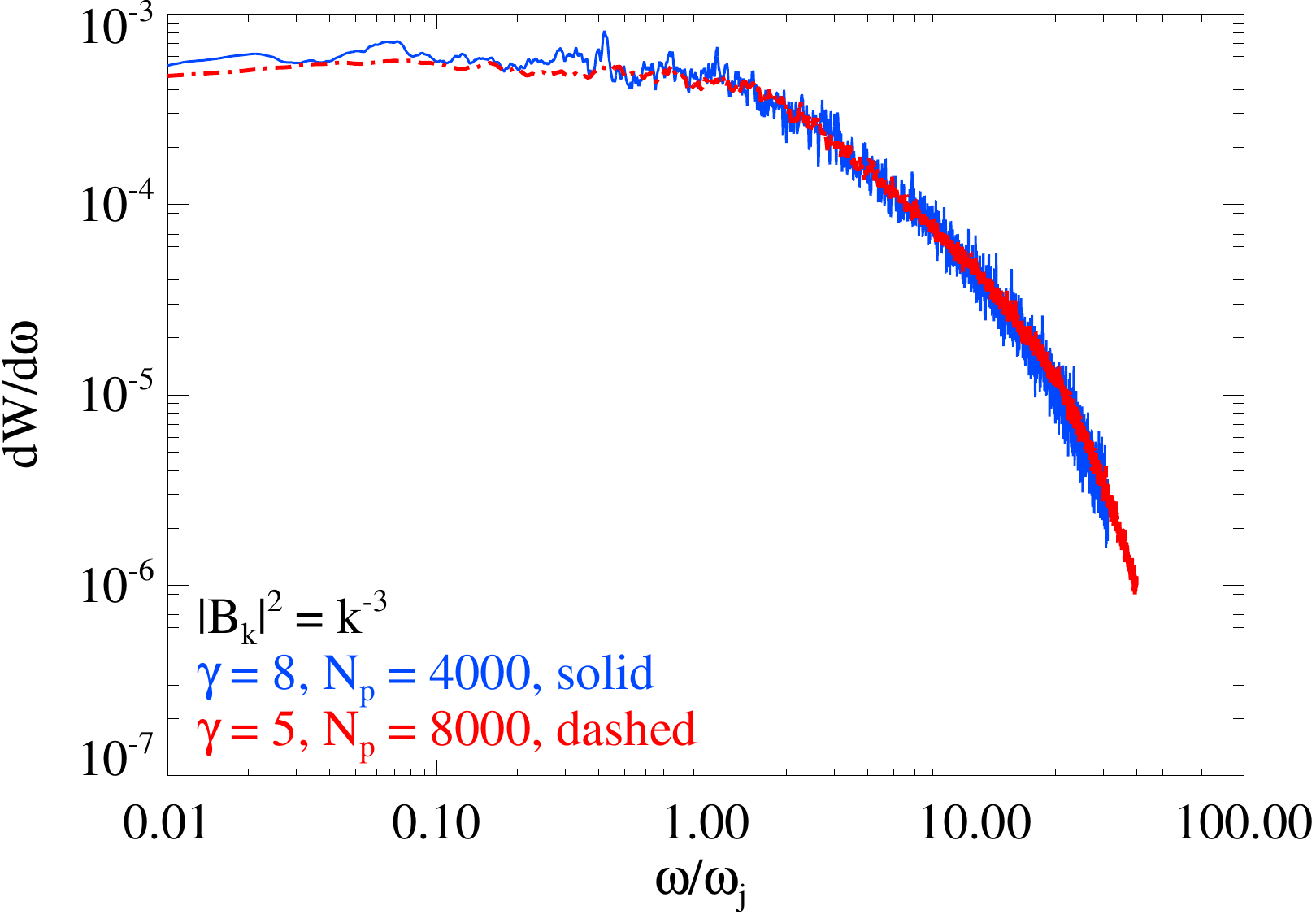}
%\vskip-0.2cm
\caption{Spectra for $\gamma$ = 5 and $\gamma = 8$ superimposed. By normalization of the frequency by $\omega_j$ = $\gamma^2k_\text{min}c$, one can clearly see the invariance of the spectral shape with respect to $\gamma$. Additionally, there is a striking contrast in the degree of ``noise'' in the two spectra. The two spectra differ in particle number by a factor of 2, with the sharpest spectrum given by the $\gamma$ = 5, with $N_p$ = 8000. }
\label{shape_comp}
\end{figure}

The spectrum also depends on the particles' energy. Figure \ref{kmax_comp} demonstrates that the break frequency $\omega_b$ scales $\propto\gamma^2$ and, moreover, the shape of the radiation spectrum is independent of the particle's Lorentz factor, as seen in Figure \ref{shape_comp}. 

\begin{figure}
\includegraphics[angle = 0, width = 1\columnwidth]{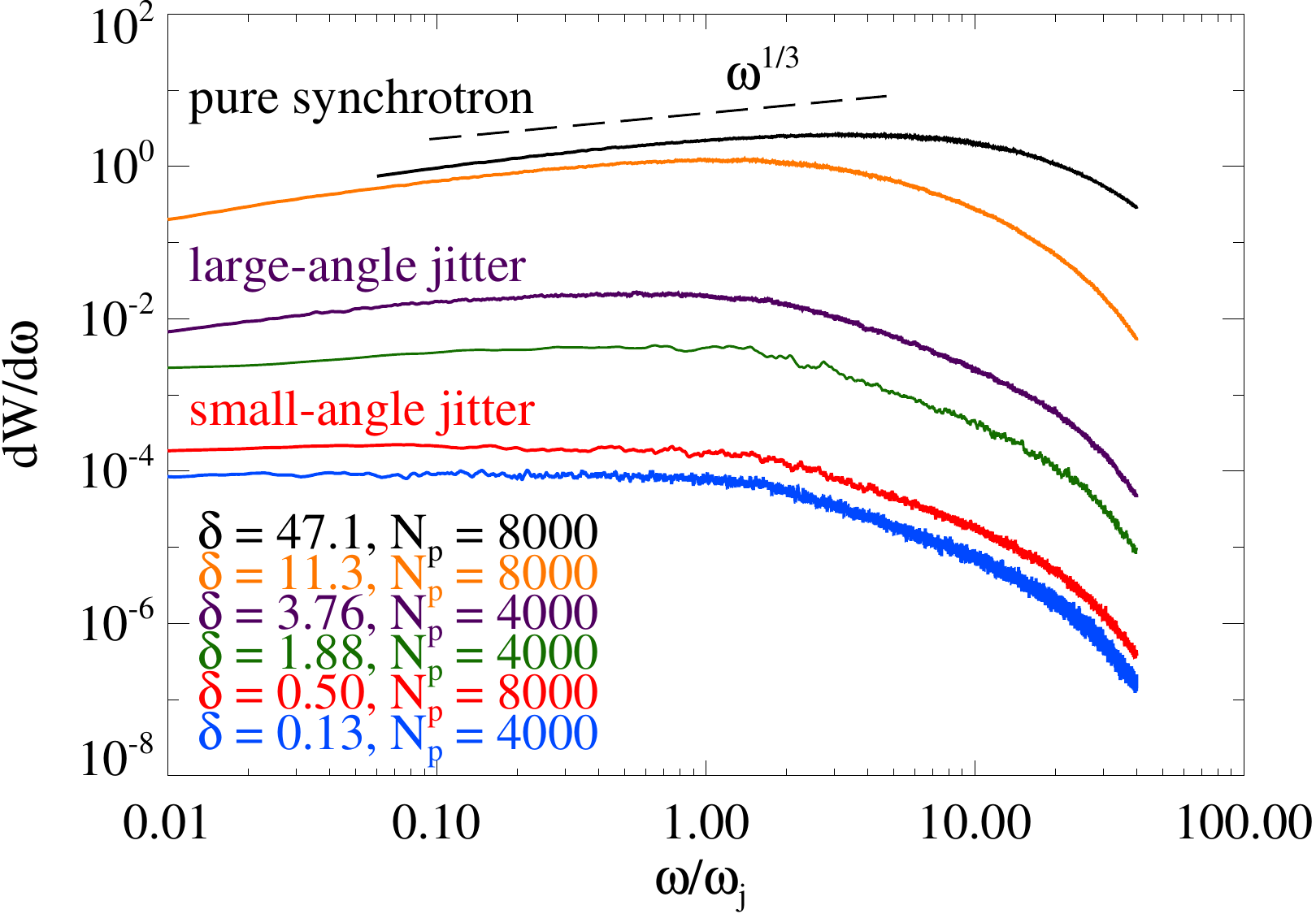}
%\vskip-0.2cm
\caption{Radiation spectra with variable jitter parameters. As can be seen, if the jitter parameter, $\delta$ is less than one -- the radiation spectrum is restricted to the small-angle jitter regime. As $\delta$ becomes greater than one, a transition occurs to the large-angle jitter regime, and then the synchrotron spectrum (with its distinctive 1/3 slope) emerges as $\delta$ becomes much greater than one. The Lorentz factor is 5 in each case.}
\label{jitter_comp}
\end{figure}

\begin{figure}
\includegraphics[angle = 0, width = 1\columnwidth]{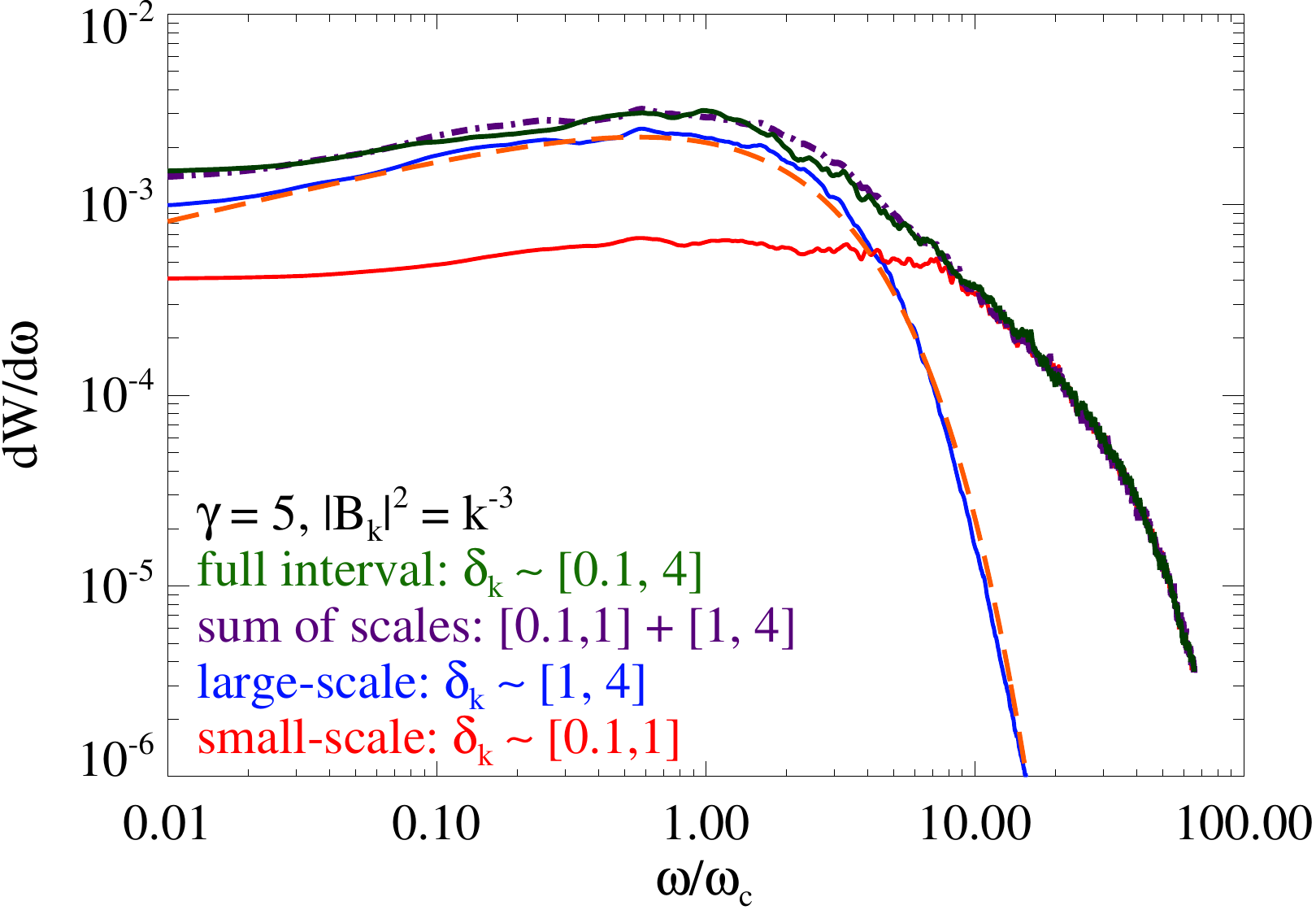}
%\vskip-0.2cm
\caption{Spectrum from the small-scale field contribution ($\delta_k \in [0.1, 1]$) superimposed with the spectrum from the large-scale field contribution ($\delta_k \in [1, 4]$), their sum, and the full scale interval ($\delta_k \in [0.1, 4]$). Each wave number of the magnetic field gives rise to its own ``scale", and therefore its own jitter parameter. The small-scale field contributes to a jitter spectrum (in red), while the large-scale field contributes to a synchotron-like spectrum (in blue). The orange dashed line indicates a typical synchrotron spectral shape. The fact that the sum of the spectra from the two scales (purple curve) entirely agrees with the spectrum from the full $\delta$ interval (green curve) illustrates the independence of radiation production at different scales.}
\label{scale_comp}
\end{figure}

Finally, the radiation spectrum depends on the jitter parameter, $\delta$ and on the regime of radiation, small-angle vs. large-angle, i.e., $\delta<1$ vs. $\delta>1$, respectively. In Figure \ref{jitter_comp}, we have superimposed spectra of increasing jitter parameter. We see that radiation is in the small-angle jitter regime for $\delta<1$, which is evident from the overall spectral shape: a flat part below the main break, a decaying power-law above it, and independence of the spectral slopes and the positions of the break on the value of $\delta$-parameter. As $\delta$ increases to values somewhat greater than unity, $1<\delta<\gamma$, the spectrum acquires some synchrotron features: the characteristic slope of $1/3$ right after the peak and the change of the peak position with $\delta$ or, equivalently, with $\langle B^2\rangle$, indicating that it is at the synchrotron frequency $\omega_c$. However, the high-frequency power-law is still well-established above the spectral peak -- in contrast to the synchrotron exponential decay at large $\omega$'s. The jitter break at $\omega_j\sim\delta^{-3}\omega_c$ (see Ref. \cite{medvedev11}) is not well seen because of a rather narrow ``dynamical range'' of the spectrum between $\omega_c$ and the fundamental frequency $\sim\omega_c/\gamma^3$ for our $\gamma=5$ runs. At even higher values of the jitter parameter, $\delta>\gamma$, the spectrum becomes purely synchrotron.  

The regime of moderately large deflections, $\delta\gtrsim1$, is interesting, as it produces a ``hybrid'' jitter-synchrotron spectrum. In this regime, $\delta_k$ of Eq. (\ref{delta-k}) exceeds unity for some $k$ and below unity for others. In this regime, the global $\delta$-parameter gives a poor description of the radiation regime, since a low-$k$ part of the magnetic field power spectrum is in the large-angle regime, whereas a high-$k$ end produces small-angle jitter radiation. To demonstrate this, we have done the following experiment. In Figure \ref{scale_comp}, we have divided a wave number range into the small-scale portion (i.e., all k for $\delta_k < 1$) and the large-scale portion (i.e., all k for $\delta_k > 1$). We see that the large-scale component produces a synchrotron-like spectrum (blue, solid curve); the dashed orange curve designates the typical synchrotron spectral shape, and the small-scale portion is in the small-angle jitter regime (red, solid curve). Additionally, we see that the sum of the two spectra (purple, dot-dashed curve) is nearly identical to the spectrum obtained from the entire wave number interval (green curve).

\section{Conclusions}
\label{s:concl}

In this paper we explored relativistic particle transport (diffusion) and radiation production in small-scale electromagnetic turbulence. We demonstrated that in the regime of small deflections, when the particle's  deflection angle is smaller than the relativistic beaming angle, $\Delta\alpha\ll1/\gamma$, the pitch-angle diffusion coefficient and the simultaneously produced radiation spectrum are tightly related, cf. Eqs. (\ref{Daa}) and (\ref{Daa-delta}). Moreover, the diffusion coefficient and the jitter parameter of turbulence, $\delta=\Delta\alpha/(1/\gamma)$, can readily be determined from the spectral information alone, cf. Eqs. (\ref{Daa-omegaj}), (\ref{Daa-spectrum}) and (\ref{delta-Pomegaj}), respectively. These theoretical results have further been confirmed with first-principles numerical simulations. 

Furthermore, we numerically explored how much information about the statistical properties of the underlying magnetic turbulence is present in the radiation spectrum produced by relativistic particles propagating though it. We have found that in the small-deflection-angle regime the relation of the radiation spectrum to the $B_k$ power spectrum is very tight. For example, for the power-law spectrum of $B$-field, Eq. (\ref{Bk}), the produced radiation spectrum is described by a double power-law (even for monoenergetic electrons) --- in clear contrast with the standard synchrotron theory. The low- and high-frequency cut-offs of the $B_k$ spectrum set the two break frequencies: the jitter frequency, $\omega_j$, being the main (low-frequency) spectral break and the high-frequency cut-off, $\omega_b$, cf. Eqs. (\ref{omegaj-kmin}) and (\ref{omegab}). Below the jitter break, $\omega<\omega_j$, the spectrum $P(\omega)$ is universal and flat. Between the breaks, $\omega_j<\omega<\omega_b$, the spectral slope is solely uniquely determined by the $B_k$ spectral slope. Above $\omega_b$, the spectrum is strongly suppressed. All these are in full agreement with theoretical predictions, Eq. (\ref{Pomega}), \citep{medvedev11}. Note that the radiation spectra are power-laws even though the particle distribution is monoenergetic. In contrast, the synchrotron spectrum exhibits an exponential fall-off above the peak frequency in this case. Note also that in isotropic turbulence, the low-frequency spectrum is universal and flat, in contrast to the synchrotron rising spectrum with the power-law index being 1/3. 

Finally, we explored how the radiation spectrum is modified when the radiation regime transits from small-angle to large-angle jitter. As $\delta$ increases above unity, the spectrum attains some synchrotron features, namely, the peak is developed and corresponds to the field-strength-dependent synchrotron frequency rather than $\omega_j$ and the low-energy slope tends to 1/3. We have demonstrated that different scales of magnetic turbulence contribute independently to radiation in different regimes: the large-scale part of $B_k$ spectrum is in the large-angle regime, $\delta_k\gtrsim1$, whereas the small-scale part is still in the small-angle regime, $\delta_k\lesssim1$. The resultant spectrum is simply the sum of the two contributions. 

To conclude, the obtained results reveal strong inter-relation of transport and radiative properties of plasmas turbulent at sub-Larmor scales. They demonstrate how spectral information can be a powerful tool to diagnose micro-turbulence in laboratory and astrophysical plasmas.

\begin{acknowledgments}
This work has been supported by the DOE grant DE-FG02-07ER54940 and the NSF grant AST-1209665.
\end{acknowledgments}

\end{document}